\documentclass[aps,prb,twocolumn,showpacs,superscriptaddress]{revtex4-2}
\usepackage{amsfonts}
\usepackage{amsmath}
\usepackage{amssymb}
\usepackage{graphicx}
\usepackage{natbib}
\usepackage{float}
\usepackage{multirow}
\usepackage{ulem}
\usepackage{gensymb}
\usepackage{dcolumn}
\usepackage[dvipsnames]{xcolor}

\usepackage[bookmarks=false]{hyperref} 
\hypersetup{colorlinks=true,
            linkcolor=blue,
            citecolor=blue,
            urlcolor=orange}
\usepackage[resetlabels,labeled]{multibib}
\newcites{App}{App Readings}
\begin{document}
\title{Magnetic properties of half metal 
from the paramagnetic phase: \\ DFT+DMFT study of exchange interactions in 
CrO$_2$}
\author{A. A. Katanin}
\affiliation{Center for Photonics and 2D Materials, Moscow Institute of Physics and Technology, Institutsky lane 9, Dolgoprudny, 141700, Moscow Region, Russia}
\affiliation{M. N. Mikheev Institute of Metal Physics of Ural Branch of Russian Academy of Sciences, S. Kovalevskaya Street 18, 620990 Yekaterinburg, Russia}

\begin{abstract}
 We study magnetic properties of CrO$_2$ within the density functional theory plus dynamical mean-field theory (DFT+DMFT) approach in the paramagnetic phase. We consider the 3-orbital (per Cr site) model, containing only $t_{2g}$ states, the $5$-orbital model, including all $d$-states, as well as the model including also the oxygen $p$-states. Using the recently proposed approach of calculation of exchange interactions in paramagnetic phase, we extract exchange interaction parameters and magnon dispersions for these models. While the magnon dispersion in the 3-orbital model possesses negative branches in accordance with previous studies in ferromagnetic phase, this drawback is removed in the $5$-orbital model. The model including oxygen states (with purely local interaction at chromium sites) overestimates the exchange interactions and spin wave stiffness. While this overestimate is partly corrected by including non-local interaction between chromium and oxygen states within the static mean-field approximation, the $5$-orbital model appears as most adequate for describing magnetic properties of CrO$_2$ with local Coulomb interaction. The possibility of describing magnetic properties of this material starting from paramagnetic phase points to the correspondence of magnetic properties in this phase and ferromagnetic phase, as well as 
important contributions of double exchange in paramagnetic phase of CrO$_2$.
\end{abstract}. 
\maketitle

\section{Introduction}
\label{sec:intro}
Half metals represent an important class of magnetic materials, see, e.g., the review \cite{Kats_rev}. Having  gapped minority spin band at the Fermi level in the ferromagnetic state, these systems can possess large magnetic moment, which finds its industrial applications. The properties of these systems are expected to be somewhat different from the strong magnets with both, minority and majority states present at the Fermi level. In the latter case large magnetic moment originates from the electron localization induced by Hund exchange \cite{Hund1,OurFe,leonov_fe} and exchange interaction is of RKKY type \cite{Vonsovsky,Stearns,AKAS1}. 

The prominent example of half metals with large magnetic moment is CrO$_2$, which has Curie temperature $T_C\simeq 390$~K and saturation magnetic moment $\mu_s\simeq 2\mu_B$ per formula unit \cite{Magn2,DFT1}. The magnetic susceptibility shows the Curie-Weiss law with the square of magnetic moment $\mu_{\rm CW}^2=(8.3\pm 0.3)\mu_B^2$ determined from the slope of inverse susceptibility\cite{Magn1,Magn2}, which also corresponds to the effective spin $S_{\rm eff}\simeq 1$, in agreement with the above mentioned saturation magnetic moment. These features can be considered as an indication of strong magnetism with well formed local magnetic moments. 

Near the Curie temperature, when the Stoner splitting is small, strong magnetic half metals are expected to reveal closer similarity to the other strong magnets. The important question is therefore whether magnetic properties of these systems originate from the presence of local magnetic moments, and whether they strongly change between the low-temperature limit and in the proximity of Curie temperature. 
The related problem is whether the effects of interaction in such strong half metal magnets are more important than pecularities of band structure yielding half metallicity. Several experimental observations (photoemission, soft-x-ray absorption and resistivity) show importance of correlation effects in CrO$_2$ \cite{Exp_Int1,Exp_Int2,Bad_metal,Exp_Int3}. Moderate correlation effects were also observed in the angle-resolved photoemission (ARPES) experiments \cite{GiorgioARPES}. The conclusions of the latter study are also supported by bulk-sensitive photoemission data,
reported in Ref. \cite{GiorgioARPES}, unveiling the occupied band structure of
CrO$_2$ in the magnetic phase.

On the theoretical side, the density functional theory (DFT) calculations of CrO$_2$ \cite{DFT1,DFT2,Bad_metal,DFT4} revealed splitting of the $d$ states into the low lying $t_{2g}$ states, which cross the Fermi level, and hybridyzed with the oxygen states, and the $e_g$ states, pushed above the Fermi level. In turn, the $t_{2g}$ states are split into the lower $xy$ state and $yz\pm xz$ excited states (the notation of the states refer to the local coordinate frame). The dispersion of the $xy$ states is almost flat, which promotes the interaction effects. In particular, the localization of the $xy$ states by the interaction effects was suggested in Ref. \cite{Korotin}. The importance of correlation effects was also emphasized in the subsequent L(S)DA+DMFT studies \cite{DMFT1,DMFT2,DMFT3,DMFT4,GiorgioARPES}. 

In accordance with the localization of the $xy$ $t_{2g}$ states and more itinerant nature of the $yz+xz$ states the double exchange nature of magnetic exchange was proposed in Refs. \cite{Korotin,Schlottmann}. Yet, recent experimental studies did not find mixed valence of chromium atoms \cite{MV2,MVnew}, in contrast to the previous results \cite{MV1}. The exchange interactions in CrO$_2$ were studied using the DFT \cite{Ex1,Solovyev}, Hartree-Fock \cite{Solovyev,Solovyev1,Solovyev_rev}, and the combination of DFT with the dynamical mean field theory (DFT+DMFT) approach \cite{Solovyev}, which produce diverse values of exchange interactions. Application of the DFT+DMFT approach to the effective 3-orbital model, containing $t_{2g}$ states only, produced negative branches of the magnon dispersion, pointing to the instability of ferromagnetism in that model \cite{Solovyev}. The authors of Ref. \cite{Solovyev} suggested inclusion of the oxygen states to stabilize the ferromagnetism. Therefore, despite long history of studying CrO$_2$, there is no common view on the mechanism of magnetic exchange and the magnitude of exchange interactions in this material. 

Recently, the DFT+DMFT approach to treat the exchange interactions in the paramagnetic state was proposed \cite{OurJq}. This approach provides a possibility to study exchange interactions without imposing certain magnetic order, which allows one to obtain an unbiased information about these interactions. 
For strong half metals, like CrO$_2$ this may also help to emphasize the effect of correlations, especially near Curie temperature, where the corresponding magnetic splitting of the states is small. 

In the present paper we revisit the problem of magnetism of CrO$_2$ within the DFT+DMFT approach. We show that in agreement with the earlier considerations the $xy$ $t_{2g}$ states appear to be more localized. We furthermore apply the recently proposed technique of calculation of exchange interactions in paramagnetic phase within the DFT+DMFT approach \cite{OurJq}. Using the obtained exchange interactions, we also obtain magnon dispersions and show that they are qualitatively and semi-quantitatively similar to those obtained in the ferromagnetic state. Remarkably, the magnon dispersion in the 5-orbital model (per chromium site) is positively definite, providing stability of ferromagnetism due to the $e_g$ states.

Therefore, on the basis of these results, we show that magnetic properties of half metals can be well described starting from the paramagnetic phase, showing the correspondence of the properties of the symmetric and symmetry broken phases of these systems. 
Among considered models, we find that the low enenrgy 5-orbital model (per Cr site), 
is quantitatively sufficient to describe ferromagnetism of CrO$_2$. We argue that 11-orbital model (per Cr site) requires considering effects of the non-local Coulomb interaction.

\section{Methods} 

\subsection{DFT}

The CrO$_2$ has P4$_2$/mnm space group (point symmetry group $D_{4h}$). The DFT calculations were performed using the pseudo-potential method implemented in the Quantum Espresso \cite{QE} package supplemented by the maximally localized Wannier projection onto $3d$ states of Cr performed within Wannier90 package \cite{Wannier90}, which produces the resulting tight-binding 5-orbital model (here and in the following we specify the number of the orbitals per Cr site, the actual number of orbitals in the respective models is doubled because of the two sites in the unit cell). For comparison, we also considered the tight-binding Hamiltonian, which includes the $p$ oxygen states, resulting in the 11-orbital model per Cr site. We use the lattice parameters $a=4.422$\AA, $c=2.916$\AA \cite{Lattice,Solovyev}.
The reciprocal space integration was performed using ${16\times 16\times 16}$\, ${\bf k}$-point grid. 

  \begin{figure}[t]
		\center{		\includegraphics[width=0.65\linewidth,angle=-90]{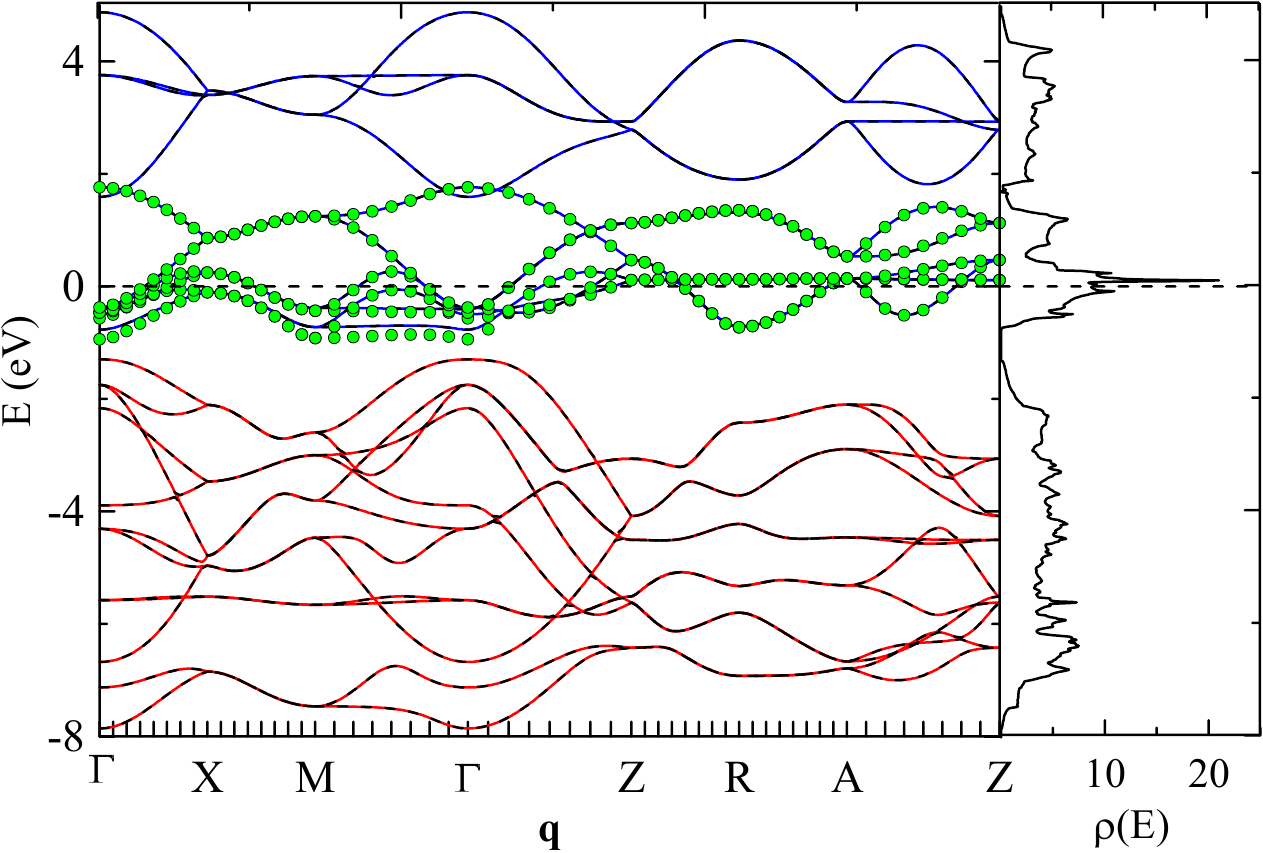}}
		\caption{Left plot: band structure (dashed lines) and its wannierization (solid lines) in 5-orbital model (per Cr site, including only $d$ states, blue lines) and 11-orbital (per Cr site, including $d$ states of chromium and $p$ states of oxygen, red and blue lines) models. The green circles show the band structure of the reduced $t_{\rm 2g}$ states model (3 orbitals per Cr site, see text). Right plot shows the respective density of states.}
\label{bands}
\end{figure}

The resulting band structure and the density of states are shown in Fig. \ref{bands}. 
The $e_g$ ($t_{2g}$) states can be constructed in the 5-orbital model by choosing the symmetric (antisymmetric) combination of $d_{xy}$ and $d_{3z^2-r^2}$ states, as well as $d_{xz}$ and $d_{yz}$ states in the global reference frame (we perform the transformation $d_{yz}\rightarrow -d_{yz}$ and $d_{xy}\rightarrow -d_{xy}$ on one of the two chromium sites); the third $t_{2g}$ state is identified with the $d_{x^2-y^2}$ state in the global reference frame, see Ref. \cite{Solovyev}. 
We choose the rotation angles between the above mentioned states to diagonalize the crystal field; the obtained angle $\theta_1$ of mixing of $d_{xy}$ and $d_{3z^2-r^2}$ states is close to $\pi/6$ and for another pair of states it is equal to $\pi/4$.  We note that Wannier functions of the $d$ states in the $5$-orbital model contain also an admixture of the oxygen states near the Fermi level (see Appendix \ref{Wan}), while in 11 orbital model the hybridization is accounted via the hopping parameters. 
To construct the model with 3 orbitals per Cr site, corresponding to considering only $t_{2g}$ states, we project out the resulting $e_g$ states in the 5-orbital model as $H_{\rm eff}=H_{t_{2g}}+H_{t_{2g},e_g}[\mu_{\rm DFT}-H_{e_g}]^{-1}H_{e_g,t_{2g}}$, where $\mu_{\rm DFT}$ is the DFT chemical potential and $H_i$ and $H_{ij}$ ($i,j=e_g,t_{2g}$) are the respective diagonal and off-diagonal blocks of the tight-binding Hamiltonian. We have verified, that the resulting Hamiltonian reproduces correctly the dispersion of the $t_{2g}$ states close to the Fermi level, see Fig. \ref{bands}.

\subsection{Dynamical mean field theory}

In DMFT calculations we consider the density-density interaction matrix, see the details in Ref. \cite{OurJq}. For the $5$-orbital and $11$-orbital models we have parameterized the interaction at the Cr sites by Slater parameters $F^0=1.99$~eV, $F^2=7.67$~eV, and $F^4=5.48$~eV, as obtained in Ref. \cite{Solovyev_rev}. For the $3$-orbital model we use the Kanamori parameterization with the interactions $U_K=2.84$~eV and $J_K=0.70$~eV, obtained in Refs. \cite{Solovyev,Solovyev_rev}. The corresponding Slater parameters (see, e.g., Supplementary Material of Ref. \cite{Sangiovanni}) $U_S=1.91$~eV, $J_S=1.17$~eV. The parameter $U_S$ is smaller than the corresponding parameter $U_S=F^0$ of the 5-orbital model due to screening of the interaction.
For 11-orbital model we use a double-counting correction ${H}_{\rm DC} = M_{\rm DC}\sum_{ir}  {n}_{ird} $ in the around mean-field form~\cite{AMF}, 
$J_S=(F^2+F^4)/14$. We have verified that the fully localized form of the double counting produces quite close results.


\subsection{Treatment of the $d$-$p$ inetraction}
\label{Udp}
Apart from the standard Coulomb repulsion in the chromium $d$-shell,
parameterized by Slatter parameters, we consider also a model including the $d$-$p$
chromium-oxygen interaction $U_{dp}$, as well as the repulsion between oxygen
$p$-states, parameterized by Kanamori parameters $U_{pp},$ $U_{pp}^{\prime}$, and $J_{pp}$, with the Hamiltonian
\begin{align}
H_{dp}&=U_{dp}\sum_{\langle ij\rangle}n_{i}n_{j}+U_{pp}\sum_{j,m}n_{jm\uparrow
}n_{jm\downarrow}\notag\\
&+\frac{U_{pp}^{\prime}-J_{pp}}{2}\sum_{j,m\neq m^{\prime
},\sigma}n_{jm\sigma}n_{jm^{\prime}\sigma}\notag\\
&+\frac{U_{pp}^{\prime}}{2}%
\sum_{j,m\neq m^{\prime},\sigma}n_{jm\sigma}n_{jm^{\prime},-\sigma},
\end{align}
where $i$ numerates chromium sites, $j$ numerates oxygen sites, $\langle
ij\rangle$ denotes nearest neighbours, $n_i=\sum_{m\sigma}{n_{im\sigma}}$, and $n_{im\sigma}=c^+_{im\sigma}c_{im\sigma}$. We treat these interactions within the static mean
field approximation, assuming approximately equal occupations of oxygen
$p$-orbitals, characterized by total occupation $\langle n_{\mathrm{O}}\rangle$ per oxygen atom, 
\begin{align}
H_{dp}^{\mathrm{MF}}&=\left[  U_{dp}z_{O}\langle n_{\mathrm{Cr}}\rangle+\widetilde{U}_{pp}\langle n_{\mathrm{O}}\rangle\right]
\sum_{j}n_{j}\notag\\
&+U_{dp}z_{\mathrm{Cr}}\langle n_{\mathrm{O}}\rangle\sum_{i}n_{i},\label{HMF}%
\end{align}
where $\widetilde{U}_{pp}=U_{pp}/(2n_{p})+(U_{pp}^{\prime}-J_{pp}%
/2)(1-1/n_{p}),$ $n_{p}=3$ is the number of $p$-orbitals, $z_{\mathrm{Cr}}=6$, $z_{\mathrm{O}%
}=3$ are the coordination (nearest neighbour) numbers of
chromium and oxygen sites, $\langle n_{\mathrm{Cr}}\rangle$ is  the respective chromium occupation of Cr per atom.  Following Ref. \cite{Held}, we subtract the
double counting contribution, which is equal to the oxygen and chromium energy
shifts in Eq. (\ref{HMF}) with the DFT occupations $\langle
n_{\mathrm{Cr}}\rangle_{0}$ and $\langle n_{\mathrm{O}}\rangle_{0}.$ The
resulting energy shifts of the chromium and oxygen states are given by \
\begin{align}
\Delta E_{\mathrm{Cr}}  & =U_{dp}z_{\mathrm{Cr}}\left[  \langle
n_{\mathrm{O}}\rangle-\langle n_{\mathrm{O}}\rangle_{0}\right]  \notag\\
&=-U_{dp}%
z_{\mathrm{Cr}}\left[  \langle n_{\mathrm{Cr}}\rangle-\langle n_{\mathrm{Cr}%
}\rangle_{0}\right]/r  ,\\
\Delta E_{\mathrm{O}}  & =U_{dp}z_{\mathrm{O}}\left[  \langle n_{\mathrm{Cr}%
}\rangle-\langle n_{\mathrm{Cr}}\rangle_{0}\right]  +\widetilde{U}_{pp}\left[
\langle n_{\mathrm{O}}\rangle-\langle n_{\mathrm{O}}\rangle_{0}\right]\notag\\
&=( U_{dp}z_{\mathrm{O}}-\widetilde{U}_{pp}/r)  \left[  \langle
n_{\mathrm{Cr}}\rangle-\langle n_{\mathrm{Cr}}\rangle_{0}\right]  ,
\end{align}
where $r=2$ is the ratio of oxygen and chromium sites in the formula unit, and
we have taken into account that the total number of electrons $\langle
n_{\mathrm{Cr}}\rangle+r\langle n_{\mathrm{O}}\rangle=\langle n_{\mathrm{Cr}%
}\rangle_{0}+r\langle n_{\mathrm{O}}\rangle_{0}$ is conserved. Finally,
absorbing the shift $\Delta E_{\mathrm{Cr}}$ at the chromium sites into the
chemical potential $\mu\rightarrow\mu+U_{dp}z_{\mathrm{Cr}}\left[  \langle
n_{\mathrm{Cr}}\rangle-\langle n_{\mathrm{Cr}}\rangle_{0}\right]/r$, we find
the energy shift of the oxygen $p$-states
\begin{equation}
\Delta E_{\mathrm{O}}'=(z_{\mathrm{O}}+z_{\mathrm{Cr}}/r)\widetilde{U}%
_{dp}\left[  \langle n_{\mathrm{Cr}}\rangle-\langle n_{\mathrm{Cr}}\rangle
_{0}\right]
\end{equation}
where $\widetilde{U}_{dp}=U_{dp}-\widetilde{U}_{pp}/(r z_{\mathrm{O}%
}+z_{\mathrm{Cr}}).$ In DMFT calculation we account for this energy shift as
(taken with the opposite sign) double counting correction of the oxygen
$p$-states. The parameter $\widetilde{U}_{dp}$ controls the energy shift of
oxygen states.
For calculations we consider the
parameters $U_{pp}=1.5$eV, $J_{pp}=0.5$eV, $U_{dp}=0.65$eV, $U_{pp}^{\prime
}=U_{pp}-2J_{pp},$ which yield $\widetilde{U}_{dp}\simeq0.6$eV. Since the screening of $d$-$d$ interaction by $p$ states is beyond the Hartree-Fock approximation, we use the same parameters of the local interaction within $d$-shell, as for the 5-orbital model.


\subsection{Exchange interactions}

To determine the exchange interactions we consider the effective Heisenberg model with the Hamiltonian $H=-(1/2)\sum_{{\bf q},rr'} J^{rr'}_{\bf q} {\mathbf S}^r_{\mathbf q} {\mathbf S}^{r'}_{-{\mathbf q}}$, {$\mathbf S^r_{\mathbf q}$ is the Fourier transform of static operators ${\mathbf S}_{ir}$},
where the orbital-summed on-site static spin operators ${\mathbf S}_{ir}=\sum_m {\mathbf S}_{irm}$ and
\begin{equation}
{\mathbf S}_{irm}=\frac{1}{2}\sum_{\sigma\sigma'\nu}c^+_{irm\sigma\nu}\mbox {\boldmath $\sigma $}_{\sigma\sigma'}c_{irm\sigma'\nu}
\end{equation}
is the electron spin operator, $\nu$ are the Matsubara frequencies, $c^+_{irm\sigma\nu}$ and $c_{irm\sigma\nu}$ are the frequency components of the electron creation and destruction operators at the site $(i,r)$, $d$-orbital $m$, and spin projection $\sigma$, and $\mbox {\boldmath $\sigma $}_{\sigma \sigma'
}$ are the Pauli matrices.

We relate exchange parameters $J_{\bf q}$ to the orbital-summed non-local static {longitudinal} susceptibility $\chi^{rr'}_{\mathbf q}=-\langle \langle S^{z,r}_{\mathbf q}|S^{z,r'}_{-{\mathbf q}}\rangle\rangle_{\omega=0}=\sum_{mm^{\prime}}\hat{\chi}_{\bf q}^{mr, m^{\prime}r'}$ (the hats stand for matrices with respect to orbital and site indexes; $\langle \langle ..|..\rangle\rangle_\omega$ is the retarded Green's function), by (see Refs. \cite{OurJq,MyCo,Fe2C})
$J_{\mathbf q}=
\chi_{\rm loc}^{-1}-\chi_{\bf q}^{-1},$
the matrix inverse 
is taken with respect to the site indexes in the unit cell. The matrix of local susceptibilities $\chi^{rr'}_{\rm loc}=-\langle \langle S^z_{ir}|S^z_{ir}\rangle\rangle_{\omega=0}\delta_{rr'}=\sum_{{  m}{  m}'} \hat{\chi}^{{  m}{  m}',r}_{\rm loc}\delta_{rr'}$ is diagonal with respect to the site indexes. 
The non-local susceptibility is determined from the Bethe-Salpeter equation using the local particle-hole irreducible vertices \cite{OurRev}, accounting also the corrections for the finite frequency box (cf. Refs.~\cite{OurJq,MyEDMFT}).  
The local irreducible vertices are extracted from the inverse Bethe-Salpeter equation applied to the local particle-hole vertex obtained within the DMFT \cite{OurRev}.

The DMFT calculations of the self-energies, non-uniform susceptibilities and exchange interactions were performed using the continuous time Quantum Monte Carlo method of the solution of impurity problem\cite{CT-QMC}, realized in the iQIST software \cite{iQIST}, see also Refs. \cite{OurJq,MyCo}.


\section{Results} 

\begin{figure}[t]
		\center{		\includegraphics[width=0.85\linewidth]{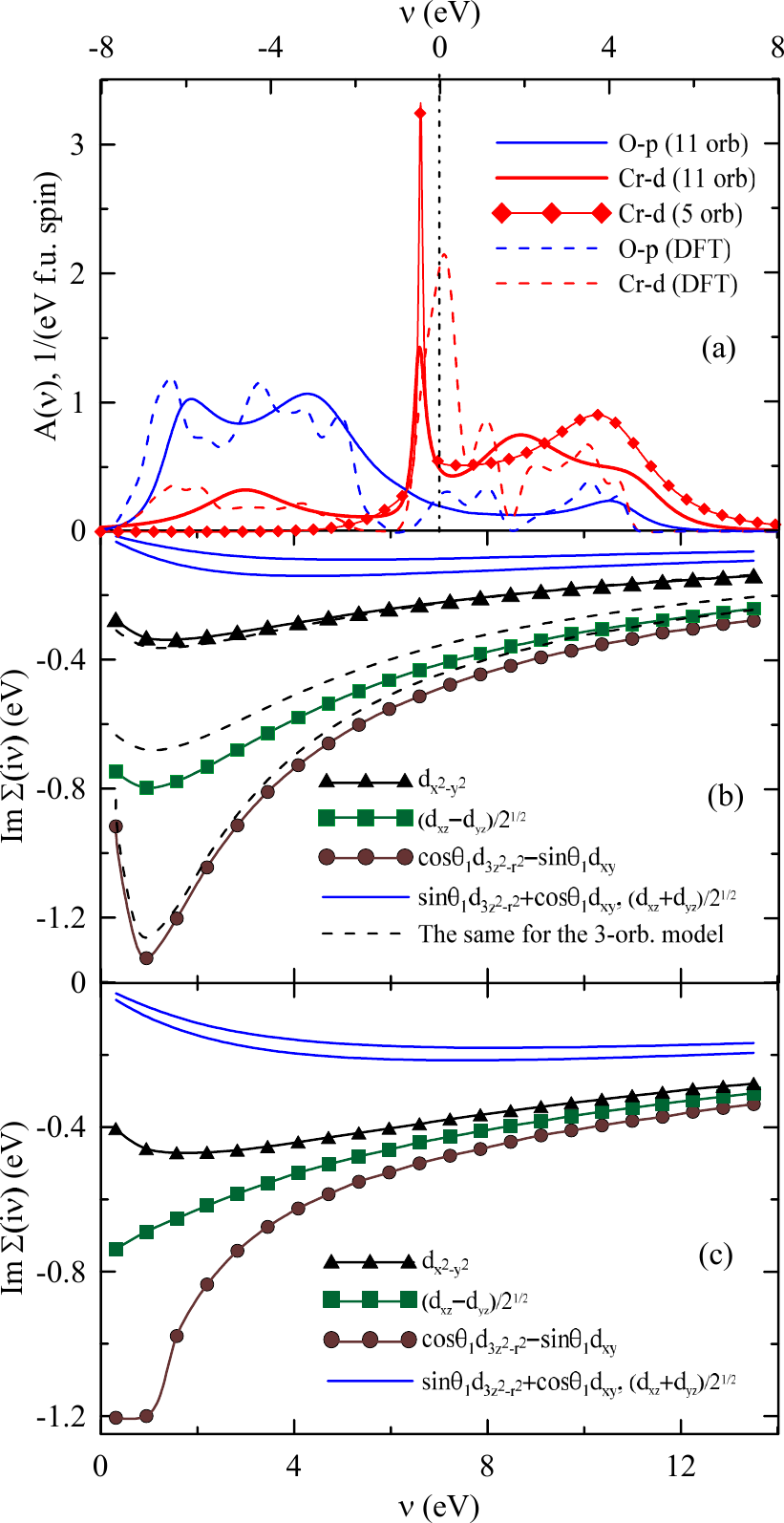}}
		\caption{
The DFT (dashed lines) and DFT+DMFT (solid lines) partial densities of states at the real frequencies (a) and the imaginary part of the self-energy at the imaginary frequency axis for states of different symmetry in five-orbital (solid lines), three-orbital (dashed lines) (b) and 11-orbital (c) model at $\beta=10$~eV$^{-1}$ in the DFT+DMFT approach with on-site Coulomb repulsion.}
\label{Fig_Sigma}
\end{figure}

In Fig. \ref{Fig_Sigma}(a) we show the partial densities of states for $\beta=10$~eV$^{-1}$, compared to those in DFT approach. The occupation of Cr sites is fixed to $2$ electrons per site in $3$- and $5$-orbital models, corresponding to Cr$^{4+}$ state of these sites. We note that this state qualitatively agrees with the substantial degree of ionicity of CrO$_2$, equal to 0.85, as estimated accoring to the recent studies \cite{Oganov}. The respective hybridized low energy oxygen and chromium states, forming Wannier functions, are considered as interacting ones. At the same time, in $11$-orbital model the occupation is determined by the total filling of $28$ electrons per unit cell, and constitutes $n_d=3.75$ per Cr site, corresponding to covalent bonding (see fillings of the $d$ orbitals in Table \ref{TabFil} and density-density correlators in Appendix \ref{DD}). The increase of the filling originates from strong hybridization of chromium and oxygen states, as discussed earlier in DFT approaches \cite{DFT1,DFT2,Bad_metal,DFT4,Korotin,Schlottmann}. We have verified that charge self-consistent DFT+DMFT treatment \cite{CSC} produces close results in view of close occupation of $d$-orbitals in DFT ($n_d=3.78$) and DFT+DMFT approach.

\begin{table}[b]
\centering
\begin{tabular}
[c]{||l||l|l|l||l|l||l||}\hline\hline
& $l_{xy}$ & $l_{xz-yz}$ & $l_{xz+yz}$ & $l_{3z^{2}-r^{2}}$ &
$l_{x^2-y^2}$ & $n_d$\\\hline\hline
3-orb & \multicolumn{1}{r}{0.49} & \multicolumn{1}{|r}{0.34} &
\multicolumn{1}{|r||}{0.17} & \multicolumn{1}{r}{0} &
\multicolumn{1}{|r||}{0} & 2\\\hline
5-orb & \multicolumn{1}{r}{0.47} & \multicolumn{1}{|r}{0.36} &
\multicolumn{1}{|r||}{0.14} & \multicolumn{1}{r}{0.02} &
\multicolumn{1}{|r||}{0.01}&2 \\\hline
11-orb & \multicolumn{1}{r}{0.52} & \multicolumn{1}{|r}{0.45} &
\multicolumn{1}{|r||}{0.39} & \multicolumn{1}{r}{0.27} &
\multicolumn{1}{|r||}{0.26} &3.75\\\hline
11-orb + $U_{dp,pp}$ & \multicolumn{1}{r}{0.48} & \multicolumn{1}{|r}{0.40} &
\multicolumn{1}{|r||}{0.25} & \multicolumn{1}{r}{0.16} &
\multicolumn{1}{|r||}{0.15} &2.87\\\hline
\end{tabular}
\caption{Fillings $n(l)$ of $d$-orbitals $l_\alpha$ per one spin projection and the total filling of $d$-states $n_d$ in DFT+DMFT approach. The notation of the orbitals refer to the local coordinate frame; the fillings are estimated at $\beta=10$~eV$^{-1}$, but only weakly depend on temperature.}
\label{TabFil}
\end{table}

\begin{figure*}[t]
		\center{		\includegraphics[width=0.8\linewidth]{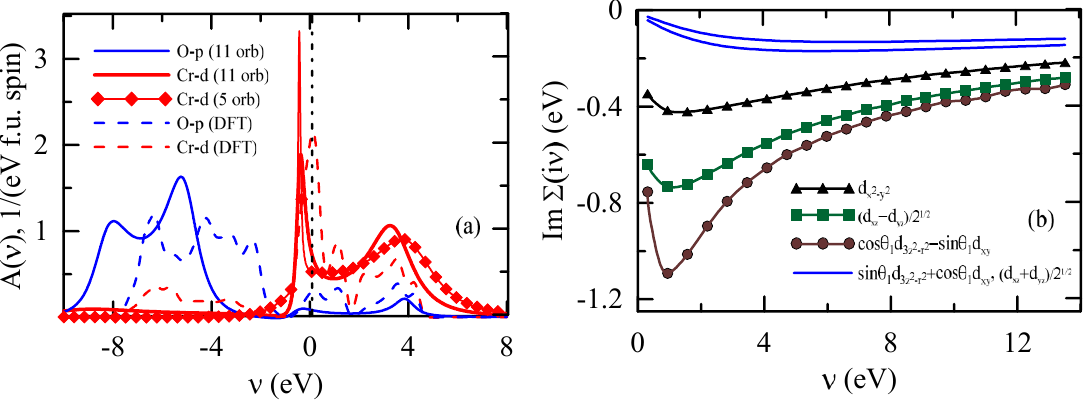}}
		\caption{Spectral functions (a) and electronic self-energies (b) in the 11-orbital model with account of $U_{pp}$ and non-local $U_{dp}$ interactions}
\label{AS11dp}
\end{figure*}


In Figs. \ref{Fig_Sigma}(b,c) we show the frequency depndence of the electronic self-energies. In agreement with previous considerations, the two of the three $t_{2g}$ states $l_{xy}=\cos\theta_1 d_{3z^2-r^2}-\sin\theta_1 d_{xy}$ and $l_{xz-yz}=(d_{xz}-d_{yz})/\sqrt{2}$ (the indexes of $l_i$ refer to the local reference frame according to Refs. \cite{Korotin,DFT4,Solovyev}) appear to have larger damping, and, respectively, more localized. On the other hand, the $t_{2g}$ state $l_{xz+yz}=d_{x^2-y^2}$, 
as well as $e_g$ states have smaller damping, and appear to be more itinerant. Closer proximity of the $t_{2g}$ states $l_{xy}$ and $l_{xz-yz}$ to half filling in the 11-orbital model  provides enhancement of correlations (cf. Ref. \cite{Toschi}), in particular non-quasiparticle form of the self-energy of these states with $\partial{\rm Im}\Sigma(i\nu)/\partial\nu=\partial{\rm Re}\Sigma(\nu)/\partial\nu>0$ at small frequencies, which yields larger local magnetic moments (see below). 

To treat properly hybridization of $d$ and $p$ states, we additionally consider the 11-orbital model, including repulsion $U_{pp}$ between oxygen $p$ states, as well as non-local interaction $U_{dp}$ between chromium $d$ and oxygen $p$ states within the static mean-field approximation, together with DMFT for the chromium $d$ states (see Sect. \ref{Udp}). The respective fillings (see Table \ref{TabFil}) in the presence of these additional interactions become closer to the 5-orbital model; the filling of chromium $d$ states constitutes in this case $2.87$ electrons.   
The results for the spectral function and self-energy of the 11-orbital model with included interactions $U_{pp}$ and $U_{dp}$ are shown in Fig. \ref{AS11dp}. One can see that the shift of $p$-states leads to suppression of $d$-$p$ hybridization, such that both, the spectral functions and self-energy become close to those in $5$-orbital model.

One can also see that, in agreement with the earlier studies \cite{DMFT4}, in all considered models the peak of the density of states, which is present in DFT approach at the Fermi level, is pushed to the energy $\nu_{\rm peak}\sim -0.5$~eV in DFT+DMFT approach.
We have verified that the $l_{xy}$ state in the considered 5- and 11-orbital models
provides largest contribution to the peak of the density of states close to the Fermi energy,
which shift can be therefore identified 
with the large quasiparticle damping of the corresponding states. This shift is therefore similar to the earlier discussed in two-dimensional systems gap formation  in the vicinity of the antiferromagnetic state \cite{AFMQS} and the Fermi surface quasi-splitting near the ferromagnetic instability \cite{KK,AritaQS}, although in the present case large damping occurs due to electronic, rather than magnetic correlations, which implies (similarly to the antiferromagnetic state) that it does not depend on the momentum (being almost local in real space).





\begin{figure}[b]
		\center{		\includegraphics[width=0.75\linewidth]{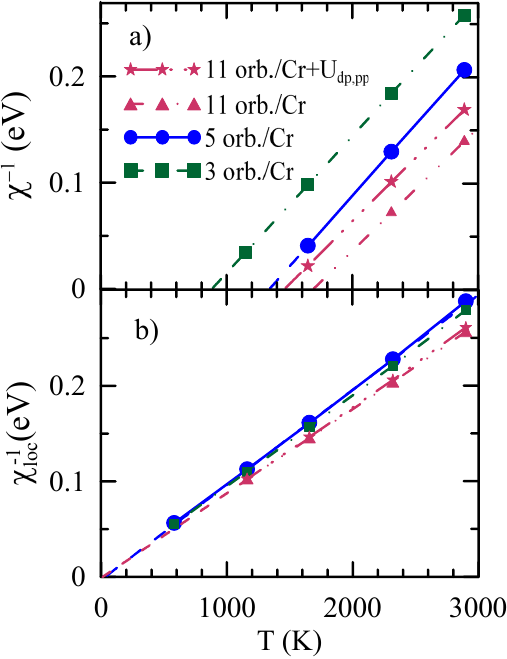}}
		\caption{
Temperature dependence of the inverse longitudinal uniform (a) and local (b) susceptibilities of CrO$_2$ within the   DFT+DMFT approach. Solid blue lines correspond to the five-orbital model per Cr site, dot-dashed green lines correspond to the three-orbital model, red dot-dot-dashed (dot-dot-dot-dashed) lines to the 11 orbital model per Cr site, including oxygen states without (with) additional $U_{dp}$ and $U_{pp}$ interactions.} 
\label{Fig_chi0_Co}
\end{figure}

 The temperature dependence of the inverse uniform $\chi_{{\bf q}=0}=\chi^{11}_{{\bf q}=0}+\chi^{12}_{{\bf q}=0}$ and local 
 \begin{equation}
 \chi_{\rm loc}=-\langle \langle S^z_{ir}|S^z_{ir}\rangle\rangle_{\omega=0}=\sum_{{  m}{  m}'} \hat{\chi}^{{  m}{  m}',r}_{\rm loc}      
 \end{equation}
 susceptibilities in the 3- and 5-orbital models, as well as 11-orbital models, is almost linear, as shown in Fig. \ref{Fig_chi0_Co}, which points to the existence of well formed local magnetic moments. The Curie temperatures, obtained from vanishing of inverse uniform susceptibility are presented in Table \ref{MomTc}.
Due to the mean-field nature the dynamical men-field theory approach is known to overestimate Curie temperature. Therefore, obtained Curie temperatures can be considered as an upper bound and corrected below with account of the non-local correlations.  

From the slope of inverse local and uniform magnetic susceptibilities we extract the local magnetic moments $\mu^2_{\rm loc}$ and $\mu^2$ according to 
\begin{align}
\chi_{\rm loc}^{-1}&=3(g\mu_B)^2 (T+T_W)/\mu_{\rm loc}^2,\\ 
\chi_{{\bf q}=0}^{-1}&=3(g\mu_B)^2 (T-T_C)/\mu^2
\end{align}
where $g=2$ (see Table \ref{MomTc}). 
In terms of the effective spin, defined by $g^2 S_{\rm eff}(S_{\rm eff}+1)=\mu_{\rm loc}^2$, this corresponds to $S_{\rm eff}=1.2$ for three- and five-orbital models and $S_{\rm eff}=1.27$ for the 11-orbital models. 
We note that the magnetic moments, especially extracted from local magnetic susceptibility, are somewhat overestimated in the considered density-density approximation, which neglects transverse components of Hund exchange, see Refs. \cite{AritaHeld,Sangiovanni}.
From the uniform susceptibility we obtain somewhat smaller magnetic moments, 
which are in a reasonable agreement with the experimental data ($\mu^2/\mu_B^2=8.3\pm 0.3$, Refs. \cite{Magn1,Magn2}). The Weiss temperature $T_W$ of the inverse local magnetic susceptibility 
appears to be quite small, showing smallness of the Kondo temperature\cite{Wilson,MyComment}.

\begin{table}[t]
\centering
\begin{tabular}
[c]{||l||l|l|l|l||}\hline\hline
& $(\mu_{\text{loc}}/\mu_{\text{B}})^{2}$ & $(\mu
/\mu_{\text{B}})^{2}$ & $T_{C}^{\text{DMFT}}$ & $T_{C}^{\text{fluct}}%
$\\\hline\hline
3-orb & \multicolumn{1}{c|}{10.6} & \multicolumn{1}{c|}{8.0} &
\multicolumn{1}{r|}{897} & \multicolumn{1}{r||}{-}\\\hline
5-orb & \multicolumn{1}{c|}{10.4} & \multicolumn{1}{c|}{7.8} &
\multicolumn{1}{r|}{1350} & \multicolumn{1}{r||}{540}\\\hline
11-orb & \multicolumn{1}{c|}{11.7} & \multicolumn{1}{c|}{8.9} &
\multicolumn{1}{r|}{1700} & \multicolumn{1}{r||}{850}\\\hline
11-orb + $U_{dp,pp}$ & \multicolumn{1}{c|}{11.4} & \multicolumn{1}{c|}{8.7} &
\multicolumn{1}{r|}{1470} & \multicolumn{1}{r||}{820}\\\hline
Experimental &
& \multicolumn{1}{c|}{8.3$\pm0.3$} & \multicolumn{1}{r|}{} &
\multicolumn{1}{r||}{390}\\\hline\hline
\end{tabular}
\caption{Magnetic moments and Curie temperatures in DFT+DMFT approach. The notation of the orbitals refer to the local coordinate frame.}
\label{MomTc}
\end{table}


\begin{table}[b]
\centering
\begin{tabular}{||c|c|c|c|c|c|c|c|c|c|c||}\hline\hline
$N_{\rm orb}$& $J_{1}$ & $J_{2}$ & $J_{3}$ & $J_{4}$ & $J_{5}$ & $J_{6}$ & $J_{7}^{>}$ &
$J_{7}^{<}$ & $J_{8}^{>}$ & $J_{8}^{<}$\\\hline
3& $11.4$ & $0.1$ & $0.6$ & $0.1$ & $-0.5$ & $-2.1$ & $-5.6$ & $-2.0$ &
$-2.0$ & $-2.0$\\\hline
5& $14.8$ & $17.8$ & $0.6$ & $0.2$ & $-0.5$ & $-1.7$ & $-5.2$ & $-1.2$ &
$-1.9$ & $-1.1$\\\hline
11 & $25.5$ & $18.1$ & $1.8$ & $0.6$ & $-1.1$ & $-2.0$ & $-5.0$ &
$-2.0$ & $-2.0$ & $-1.4$\\\hline
11$dp$ & $8.8$ & $18.0$ & $-0.1$ & $0.1$ & $-0.4$ & $-0.6$ & $-2.9$ & $-0.7$ &
$-0.7$ & $-1.1$\\\hline\hline
\end{tabular}
\caption{Exchange interactions (in meV) between various chromium sites at $\beta=10$~eV$^{-1}$ for the 3- and 5 orbital models per chromium site, as well as 11-orbital models, including oxygen states. $11dp$ stands for the model with $U_{dp}$, $U_{pp}$ interactions. The notation of the exchange interactions is the same as in Refs. \cite{Solovyev,Solovyev_rev}.}
\label{TabEx}
\end{table}

Using the approach of Refs. \cite{OurJq,MyCo,Fe2C} we obtain the exchange interactions $J^{rr'}_{\bf q}$. 
The Fourier transformation of the obtained exchange interactions at $\beta=10$~eV$^{-1}$ is presented in Table \ref{TabEx}. The obtained exchange interactions are comparable to those obtained in the ferromagnetic state in Refs. \cite{Solovyev,Solovyev_rev}, with the exchange interactions between the nearest neighbour sites larger in the presence of the oxygen states, than in the 3- and 5-orbital models, due to larger hybridization of chromium states. We note that in the 5-orbital model the magnetic exchange is mediated by hybridyzed $d$ and $p$ states constituting Wannier orbitals (see Appendix A), having also relatively small bandwidth, comparable to Hund exchange. This results in the important contribution of the double exchange in this model. At the same time, uncorerrelated oxygen states with rather large bandwidth in 11-orbital model mediate magnetic exchange of purely RKKY type. With including $U_{dp}$ and $U_{pp}$ interactions, the nearest neighbour exchange is suppressed in $11$-orbital model; the antiferromagnetic exchanges at longer distances are however also suppressed, reducing frustration effects.  

  \begin{figure}[t]
		\center{		\includegraphics[width=0.95\linewidth]{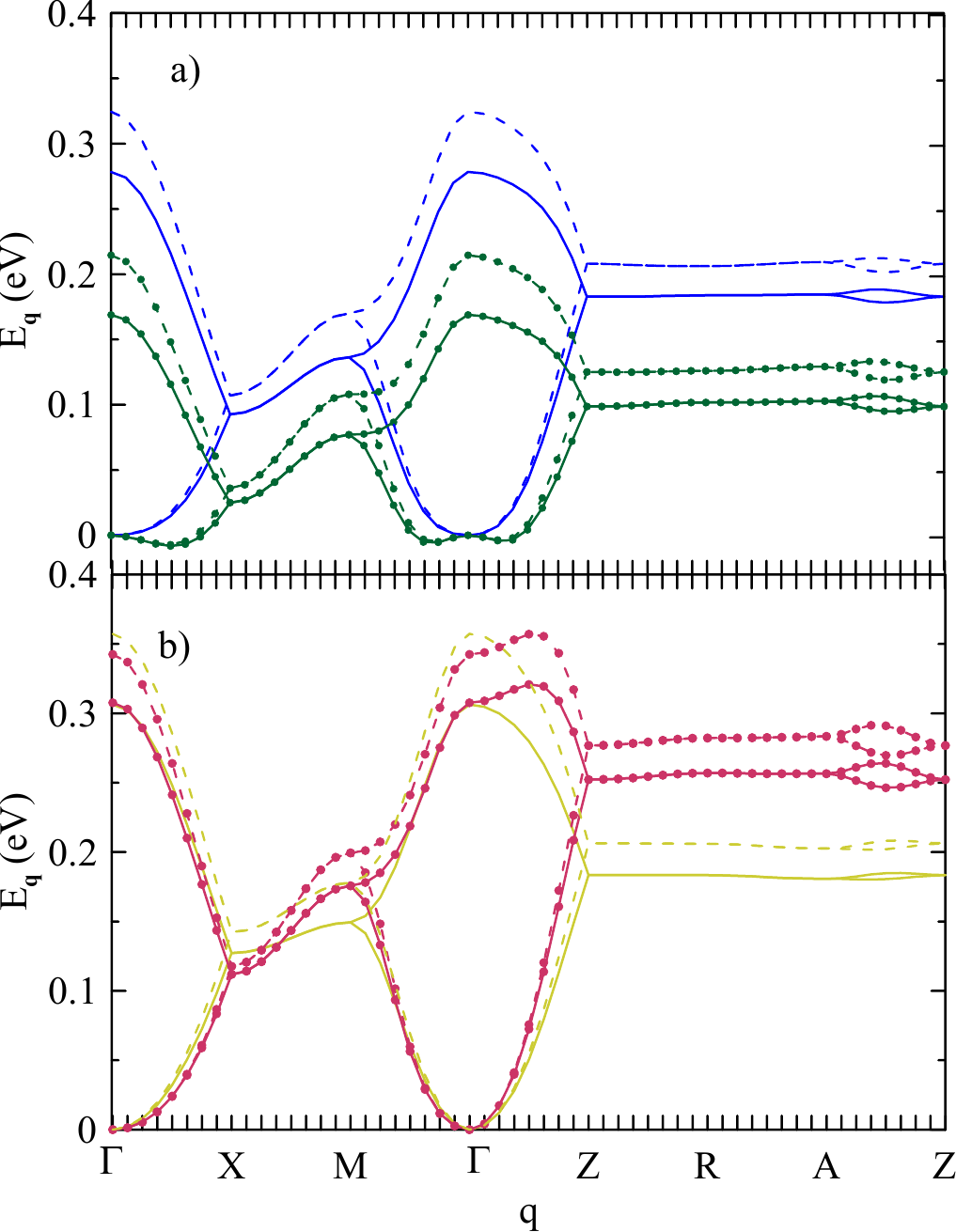}}
		\caption{Magnon dispersion at $\beta=7$~eV$^{-1}$ (solid lines) and $\beta=10$~eV$^{-1}$ (dashed lines) (a) in the models with three (green lines with symbols) and five orbitals (blue lines) per Cr site and (b) in the models with 11 orbitals per Cr site which includes oxygen states, with (dark yellow lines) or without (red lines with symbols) additional $U_{dp} $ and $U_{pp}$ interactions. }
\label{Disp}
\end{figure}

Using the obtained exchange interactions in the temperature range $T\gtrsim T_C$, we obtain magnon dispersion as the ${\bf q}$-dependent eigenvalues of the matrix of the spin-wave Hamiltonian (cf. Refs. \cite{MyCo,Fe2C}), assuming that the exchange interactions do not change strongly with lowering the temperature. 
The resulting magnon dispersions 
are shown in Fig. \ref{Disp}.  One can see that the magnon dispersion of the $3$-orbital model possesses negative branches, showing an instability of ferromagnetism, similarly to previous study in the ferromagnetic phase \cite{Solovyev}. At the same time, the magnon dispersions of the $5$-orbital model are positive definite, providing the stability of the ferromagnetic state. Therefore, inclusion of the $e_g$ states seems to be crucial for the stability of ferromagnetism. The maximal energy of the obtained magnon dispersion in the 5-orbital model is larger than that in the ``method $\hat{b}$" of Ref. \cite{Solovyev_rev}, corresponding to the infinitesimal rotation of exchange-correlation potential, but comparable to that obtained in the ``method $\hat{m}$" of Ref. \cite{Solovyev_rev} (considering infinitesimal rotation of magnetization).  The dispersion in the model including oxygen states without additional $U_{dp}$ and $U_{pp}$ interactions is somewhat larger than in the 5-orbital model  due to larger exchange interactions, but becomes comparable to that for the 5-orbital model with account of $U_{dp,pp}$ interactions.

  \begin{figure}[t]
		\center{		\includegraphics[width=0.95\linewidth]{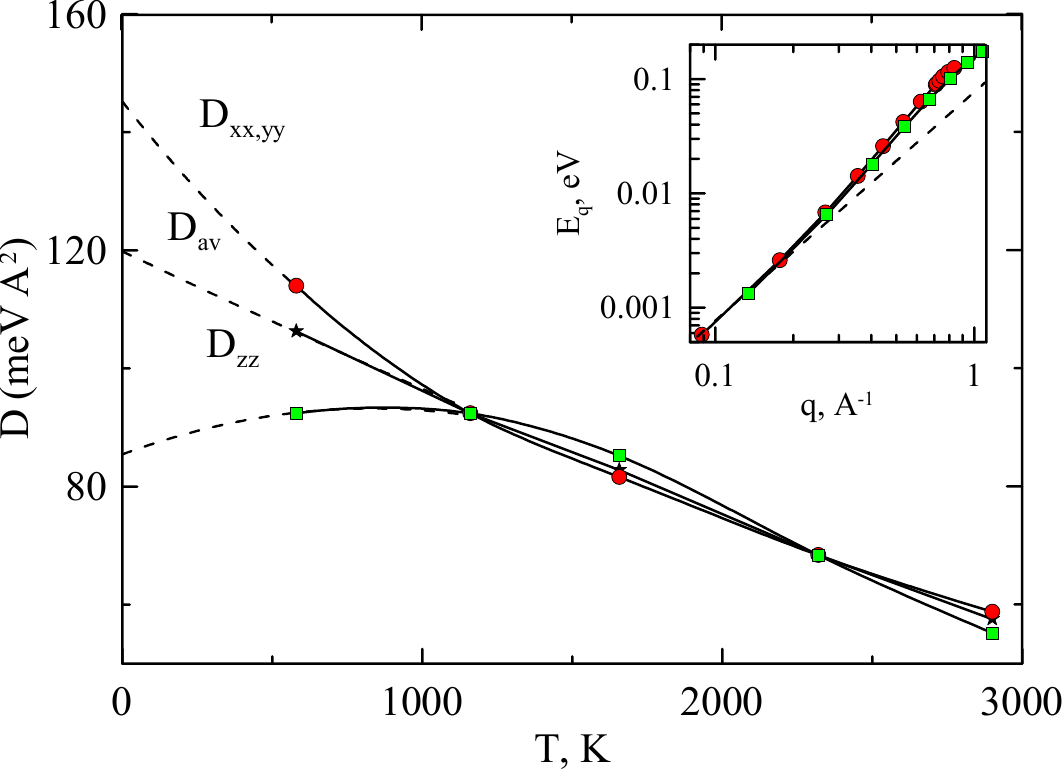}}
		\caption{Temperature dependencies of the obtained spin-wave stiffnesses in the $x,y$ (red circles) and $z$ (green squares) directions, together with the average spin-wave stiffness $D_{\rm av}=(D^2_{xx}D_{zz})^{1/3}$ (black stars) in the five-orbital model. Dashed lines show the result of extrapolation. The inset shows momentum dependencies of magnon energies at $\beta=10$~eV$^{-1}$ in the respective directions in logarithmic scale, dashed line corresponds to quadratic fit.} 
\label{D}
\end{figure}

The temperature dependencies of the obtained spin-wave stiffnesses in the 5-orbital model in various directions are shown in Fig. \ref{D}. One can see that the average spin-wave stiffness, extrapolated to the low-temperature limit, $D_{\rm av}\simeq 120$~meV$\cdot$\AA$^2$ is in a reasonable agreement with the experimental data $D=60$ to $150$~meV$\cdot$\AA$^2$, Refs. \cite{ExpFMR,ExpFilm,ExpD}. At the same time, the 11-orbital models yield larger value of the spin-wave stiffness, $D_{\mathrm av}\gtrsim 200$~meV$\cdot$\AA$^2$ (not shown). 
Although the interactions $U_{dp,pp}$ yield the suppression of exchange interactions and the spin wave stiffness $D_z$ in 11-orbital model, spin wave stiffness $D_{xy}$ is increased by these interactions due to suppression of antiferromagnetic exchange interactions $J_{7,8}$ (the suppressed ferromagnetic interaction $J_1$ acts along the $z$ axis and therefore contributes to $D_z$ only). This shows, that for accurate description of the low-energy magnon dispersion in 11-orbital model the treatment of the non-local interaction beyond simplest static mean field approximation is required.


Finally, to estimate the non-local corrections to the Curie temperature beyond DMFT, we use the RPA approach \cite{RPA_TC}, see also Ref. \cite{MyCo}. Assuming that the sites of the unit cell are equivalent, we find 
\begin{equation}
T_C=
\frac{\mu^2}
{3(g\mu_B)^2\sum\limits_{{\mathbf q}} \left[\lambda \delta_{rr'}-J_{\mathbf q}^{rr'}\right]_{11}^{-1}},
\end{equation}
where 
$\lambda=\sum_{r'} J^{rr'}_0$. The obtained results taking the obtained exchange interactions at $\beta=10$~eV$^{-1}$ are presented in Table \ref{MomTc}. With account of the non-local corrections, the Curie temperature is suppressed with respect to the DMFT Curie temperature, and for the 5-orbital model only moderately overestimates experimental data. For the 11-orbital models the Curie temperature is stronger overestimated; for the model including $U_{dp,pp}$ interactions the suppression of Curie temperature with respect to DMFT appears not too strong because of weakened frustration effects in this model. 

The success of the 5-orbital model in description of the magnetic properties of CrO$_2$ relies on the fact that this model is better suited to describe double exchange interaction, having also lower band width, comparable to Hund exchange interaction. Describing the double exchange interaction within 11-orbital model requires accurate (possibly, non-perturbative) treatment of the non-local Coulomb interactions.


\section{Conclusion}
In summary, we have evaluated non-uniform susceptibilities, Curie temperatures, and exchange interactions in 3-, 5-, and 11-orbital (per Cr site) models within DFT+DMFT approach. 
The most reasonable results are obtained within the low-energy 5-orbital model, representing double exchange interaction.
This model yields positive magnon dispersions and reasonable Curie temperature, although the latter is still overestimated with respect to the experimental data. The overestimate of the Curie temperature is likely connected with the assumed density-density form of the Coulomb interaction (cf. Ref. \cite{Sangiovanni}), presence of magnetic frustrations, etc. We show also that the considered approach allows for a correct description of the experimental data for the spin-wave stiffness. 

At the same time, the 11-orbital model, including oxygen states, yields strong hybridization of these states with chromium states at the energies well below the Fermi level, which results in the filling of d-orbitals of Cr closer to half filling, and therefore stronger correlations. Remarkably, we find RKKY mechanism of magnetic exchange, represented by 11-orbital model with local Coulomb interaction, inapplicable even in paramagnetic phase of CrO$_2$.
We argue that considering non-local interaction between chromium and oxygen sites (together with $U_{pp}$ interaction) within static mean field approximation increases occupation of oxygen $p$ states and substantially improves the results for the 11-orbital model. 
At the same time, it yields larger spin-wave stiffness, than that for 5-orbital model and experimental data. Likely, treatment of non-local interactions beyond static mean-field approximation, e.g. within cluster methods or non-local extensions of dynamical mean-field theory \cite{OurRev,MyEDMFT,Stepanov1,Stepanov2,HeldDGA1}, will further improve the results of this model. 




The possibility of describing reasonably well magnetic properties of CrO$_2$ from the paramagnetic phase 
implies presence of the correspondence between the magnetic properties in ferro- and paramagnetic phases. Mathematically, this correspondence occurs due to compensation of the self-energy and vertex corrections to the spin susceptibility, which was discussed earlier in the ferro- \cite{EdwHrtz} and paramagnetic  \cite{OurJq} phases.  
The performed study also implies formation of local magnetic moments in CrO$_2$ due to Hund exchange interaction, and their double exchange-like interaction even in paramagnetic phase.

Further experimental and theoretical studies of the form of magnon dispersion, and its evolution from the low- to the high-temperature limit are of certain interest. Also describing the effect of the non-local chromium-oxygen interaction beyond static mean field approximation, as well as the consideration of magnetic properties of CrO$_2$ with full SU(2) symmetric Coulomb interaction has to be performed in future.

\vspace{-0.2cm}
\section*{Acknowledgements} 
\vspace{-0.2cm}
The author appreciates stimulating discussions with I. V. Solovyev at the early stage of the work. The calculations were performed on the cluster of the Laboratory of Material Computer Design of MIPT.
The DMFT calculations are  supported by the project of Russian Science Foundation 19-72-30043-P. The DFT calculations are performed within the theme ``Quant" 122021000038-7
of {the} Ministry of Science and Higher Education of the Russian Federation. 

\begin{widetext}
\begin{appendix}
\numberwithin{equation}{section}

\section{Wannier orbitals in 5- and 11-orbital models}
\label{Wan}
In Figs. \ref{Wan1} and \ref{Wan2} we visualize \cite{VESTA} Wannier orbitals in 5- and 11-orbital models (before performing basis rotation which diagonalises crystal field). One can see that in the 5-orbital model Wannier functions contain substantial admixture of the oxygen states, originating from the bands the vicinity of the Fermi level, while in 11-orbital model Wannier functions are more localized at chromium and oxygen sites. As it is discussed in the main text in the latter model however the hybridization occurs via the hopping parameters, which in particular yield additional contribution to the density of $d$ states at the energies well below the Fermi level ($\nu\sim -4$~eV).
\vspace{-0.3cm}
  \begin{figure}[h!]
		\center{\includegraphics[width=0.7\linewidth]{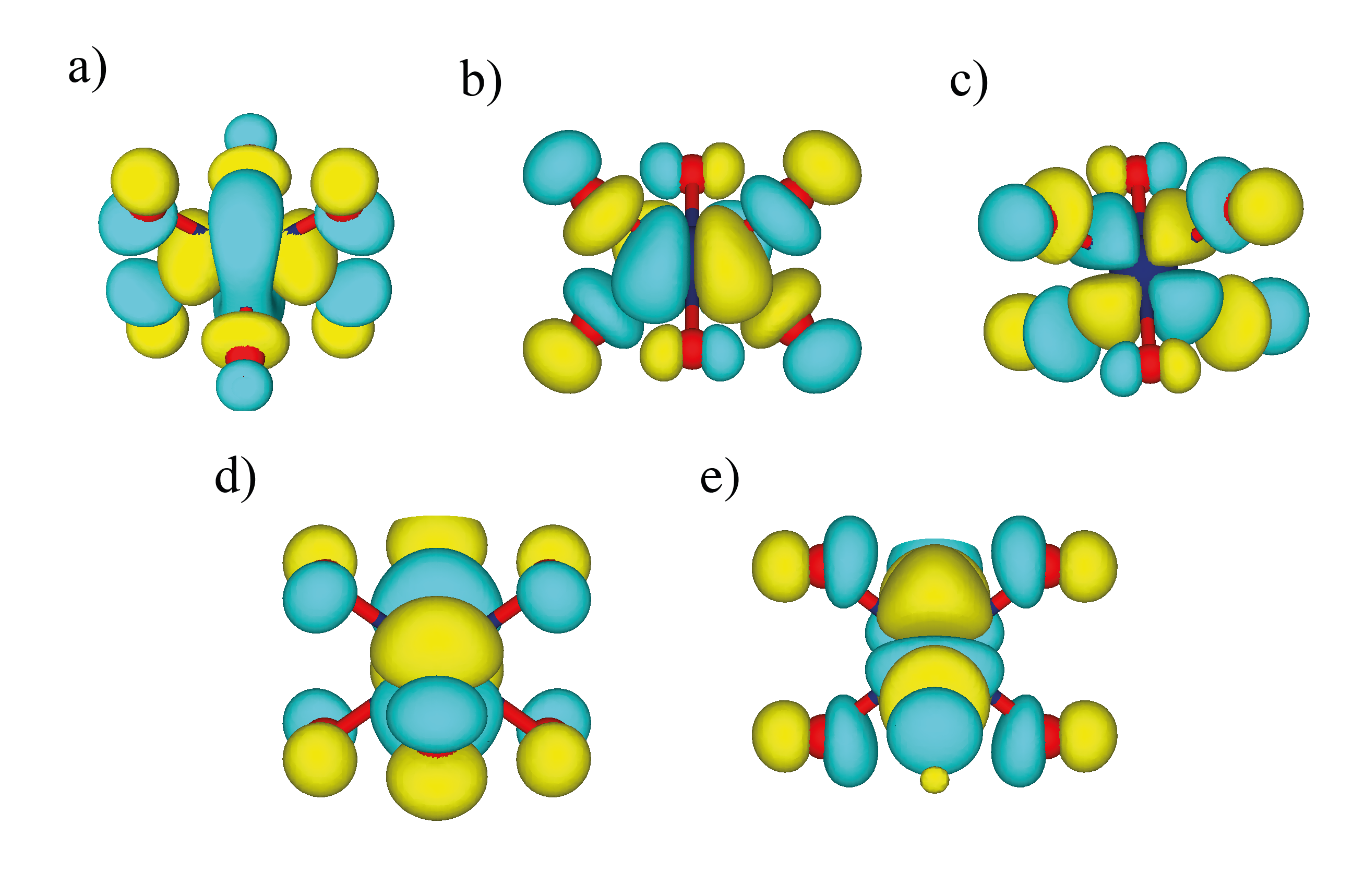}}
\caption{Visualization of Wannier orbitals in 5-orbital model (per Cr site), including only $d$ states. Blue circles in the center (partly hidden by Wannier orbitals) correspond to chromium atom, red lines and circles show the bonds and oxygen atoms.}
\label{Wan1}
\end{figure}

  \begin{figure}[t]
\center{\includegraphics[width=0.8\linewidth]{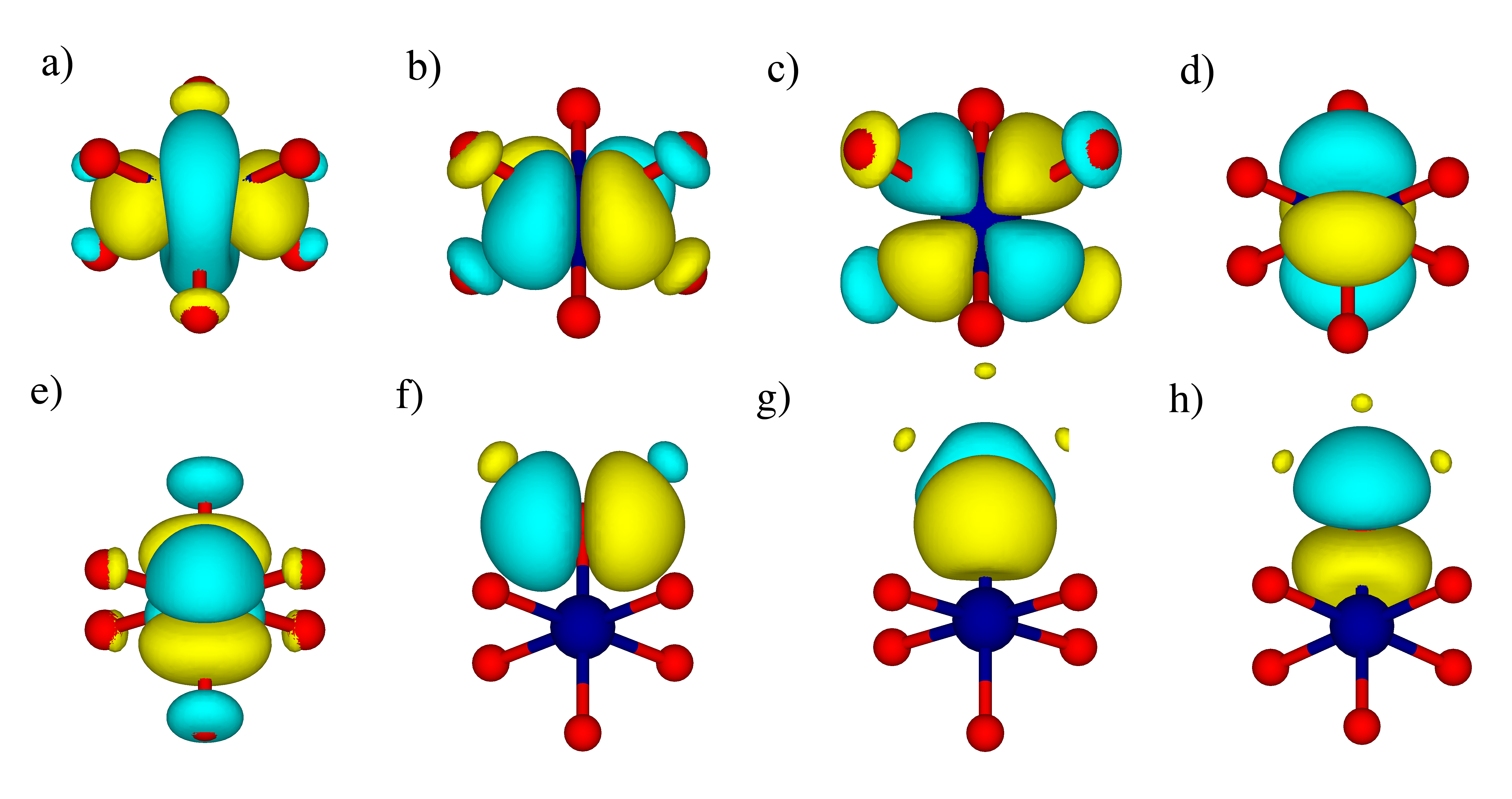}}
		\caption{Visulaization of Wannier orbitals in 11-orbital model (per Cr site) at chromium (a-e) and one of the oxygen (f-h) sites. The notations are the same as in Fig. \ref{Wan1}.}
\label{Wan2}
\end{figure}


 \section{Density correlations in DFT+DMFT approach}
 \label{DD}


In Tables \ref{DD1}, \ref{DD2}, and \ref{DD3} we present the density-density correlation function $\langle n_{m\sigma}n_{m'\sigma'}\rangle$ in 5- and 11-orbital models in DFT+DMFT approach at $\beta=10$~eV$^{-1}$. In the 5-orbital model the density correlation matrix has only few off-diagonal elements, the major one is between $l_{xy}$ and $l_{xz-yz}$ orbitals in the local coordinate frame, which have largest quasiparticle damping (see Fig. 2 of the main text). The 11-orbital model with local interaction possesses stronger off-diagonal correlations, which reflects stronger mixing of various orbital states in this model. On the other hand, in 11-orbital model with $U_{dp}$ and $U_{pp}$ interactions the correlations become more diagonal, and resemble the results for the 5-orbital model.

\begin{table*}[h!]
\begin{tabular}
[c]{|l|l|l|l|l|l|l|l|l|l|l|l|l|}\hline
  & $n_{1\uparrow}$   &  $n_{2\uparrow}$   &  $n_{3\uparrow}$   &  $n_{4\uparrow}$   & $n_{5\uparrow}$   &  $n_{1\downarrow}$   &  $n_{2\downarrow}$   &  $n_{3\downarrow}$   &  $n_{4\downarrow}$   &  $n_{5\downarrow}$   \\\hline
  $n_{1\uparrow}$ &  {\bf 0.47} &  0.01   &  0.00   &  0.33   & 0.11   &  0.00   &  0.01   &  0.00   &  0.01   &  0.01   \\\hline
  $n_{2\uparrow}$ &0.01   &  {\bf 0.02}  &  0.00   &  0.01   &
  0.00   &  0.01   &  0.00   &  0.00   &  0.00   &  0.00   \\\hline
  $n_{3\uparrow}$ &0.00   &  0.00   &    {\bf 0.01} &  0.00   &
  0.00   &  0.00   &  0.00   &  0.00   &  0.00   &  0.00   \\\hline
  $n_{4\uparrow}$ &0.33   &  0.01   &  0.00   &   {\bf 0.36} &  0.07   &  0.01   &  0.00   &  0.00   &  0.00   &  0.01   \\\hline
  $n_{5\uparrow}$ &0.11   &  0.00   &  0.00   &  0.07   &   {\bf 0.14}   &  0.01   &  0.00   &  0.00   &  0.01   &  0.00   \\\hline
  $n_{1\downarrow}$ &0.00   &  0.01   &  0.00   &  0.01   &  0.01   &  {\bf 0.47}  &  0.01  &  0.00   &  0.33   &  0.11   \\\hline
  $n_{2\downarrow}$&0.01   &  0.00   & 0.00   &  0.00   &  0.00   &  0.01   &  {\bf 0.02}  &  0.00   &
  0.01   &  0.00   \\\hline
  $n_{3\downarrow}$&0.00   &  0.00   &  0.00   &  0.00   &  0.00   &  0.00   &  0.00   &  {\bf 0.01}  &  0.00   &  0.00   \\\hline
  $n_{4\downarrow}$&0.01   &  0.00   &  0.00   &  0.00   &
  0.01   &  0.33   &  0.01   &  0.00   &  {\bf 0.36}  & 0.07   \\\hline
  $n_{5\downarrow}$&0.01   &  0.00   &  0.00   &  0.01   &  0.00   &  0.11   &  0.00   &  0.00   &
  0.07   &   {\bf 0.14} \\\hline
\end{tabular}
\caption{Density correlators $\langle n_{m\sigma}n_{m'\sigma'}\rangle$ in the 5-orbital model in DFT+DMFT approach at $\beta=10$~eV$^{-1}$. The diagonal elements correspond to the respective fillings. The notation of the orbitals: 1: $l_{xy}$, 2: $l_{3z^2-r^2}$, 3: $l_{x^2-y^2}$, 4: $l_{xz-yz}$, 5: $l_{xz+yz}$ }
\label{DD1}
\end{table*}

\begin{table*}[h!]
\begin{tabular}
[c]{|l|l|l|l|l|l|l|l|l|l|l|l|l|}\hline
  & $n_{1\uparrow}$   &  $n_{2\uparrow}$   &  $n_{3\uparrow}$   &  $n_{4\uparrow}$   & $n_{5\uparrow}$   &  $n_{1\downarrow}$   &  $n_{2\downarrow}$   &  $n_{3\downarrow}$   &  $n_{4\downarrow}$   &  $n_{5\downarrow}$   \\\hline
  $n_{1\uparrow}$ & {\bf 0.52}  &  0.16   &  0.14   &  0.35   & 0.28   &  0.08   &  0.12   &  0.12   &  0.11   &  0.12   \\\hline
  $n_{2\uparrow}$ &0.16   &  {\bf 0.27} &  0.07   &  0.12   &
  0.11   &  0.12   &  0.06   &  0.06   &  0.10   &  0.09   \\\hline
  $n_{3\uparrow}$ &0.14   &  0.07   &   {\bf 0.26}&  0.12   &
  0.10   &  0.12   &  0.06   &  0.05   &  0.10   &  0.09   \\\hline
  $n_{4\uparrow}$ &0.35   &  0.13   &  0.12   &  {\bf 0.45}  &  0.23   &  0.11   &  0.10   &  0.10   &  0.09   &  0.11   \\\hline
  $n_{5\uparrow}$ &0.28   &  0.11   &  0.10   &  0.23   &  {\bf 0.39}    &  0.12   &  0.09   &  0.09   &  0.11   &  0.08   \\\hline
  $n_{1\downarrow}$ &0.08   &  0.12   &  0.12   &  0.11   &  0.12   &  {\bf 0.52}  &  0.15   &  0.14   &  0.35   &  0.28   \\\hline
  $n_{2\downarrow}$&0.12   &  0.06   & 0.06   &  0.10   &  0.09   &  0.15   &   {\bf 0.27} &  0.07   &
  0.13   &  0.11   \\\hline
  $n_{3\downarrow}$&0.12   &  0.06   &  0.05   &  0.10   &  0.09   &  0.14   &  0.07   &  {\bf 0.26}  &  0.12   &  0.10   \\\hline
  $n_{4\downarrow}$&0.11   &  0.10   &  0.10   &  0.09   &
  0.11   &  0.35   &  0.13   &  0.12   &   {\bf 0.45} & 0.23   \\\hline
  $n_{5\downarrow}$&0.12   &  0.09   &  0.09   &  0.11   &  0.08   &  0.28   &  0.11   &  0.10   &
  0.23   &   {\bf 0.39}  \\\hline
\end{tabular}
\caption{Density correlators in the 11-orbital model with local interaction in DFT+DMFT approach at $\beta=10$~eV$^{-1}$. The notation of the orbitals is the same as in the Table \ref{DD1}.}
\label{DD2}
\end{table*}

\begin{table*}[h!]
\begin{tabular}
[c]{|l|l|l|l|l|l|l|l|l|l|l|l|l|}\hline
  & $n_{1\uparrow}$   &  $n_{2\uparrow}$   &  $n_{3\uparrow}$   &  $n_{4\uparrow}$   & $n_{5\uparrow}$   &  $n_{1\downarrow}$   &  $n_{2\downarrow}$   &  $n_{3\downarrow}$   &  $n_{4\downarrow}$   &  $n_{5\downarrow}$   \\\hline
  $n_{1\uparrow}$ &  {\bf 0.48} &  0.09&  0.07   &  0.33   & 0.18   &  0.04   &  0.07   &  0.06   &  0.05   &  0.05   \\\hline
  $n_{2\uparrow}$ &0.09   &  {\bf 0.16}  &  0.02   &  0.07   &
  0.04   &  0.07   &  0.02   &  0.02   &  0.05   &  0.03   \\\hline
  $n_{3\uparrow}$ &0.07   &  0.02   &    {\bf 0.15} &  0.06   &
  0.04   &  0.06   &  0.02   &  0.02   &  0.05   &  0.03   \\\hline
  $n_{4\uparrow}$ &0.33   &  0.07   &  0.06   &   {\bf 0.40} &  0.14   &  0.05   &  0.05   &  0.05   &  0.04   &  0.04   \\\hline
  $n_{5\uparrow}$ &0.18   &  0.04   &  0.04   &  0.14   &   {\bf 0.25}   &  0.05   &  0.03   &  0.03   &  0.04   &  0.02   \\\hline
  $n_{1\downarrow}$ &0.04   &  0.07   &  0.06   &  0.05   &  0.05   &  {\bf 0.48}  &  0.09  &  0.08   &  0.33   &  0.18   \\\hline
  $n_{2\downarrow}$&0.07   &  0.02   & 0.02   &  0.05   &  0.03   &  0.09   &  {\bf 0.16}  &  0.02   &
  0.07   &  0.04   \\\hline
  $n_{3\downarrow}$&0.06   &  0.02   &  0.02   &  0.05   &  0.03   &  0.08   &  0.02   &  {\bf 0.15}  &  0.06   &  0.03   \\\hline
  $n_{4\downarrow}$&0.05   &  0.05&  0.05   &  0.04   &
  0.04   &  0.33   &  0.07   &  0.06   &  {\bf 0.40}  & 0.14   \\\hline
  $n_{5\downarrow}$&0.05   &  0.03   &  0.03   &  0.04   &  0.02   &  0.18   &  0.04   &  0.03   &
  0.14   &   {\bf 0.25} \\\hline
\end{tabular}
\caption{Density correlators in the 11-orbital model in DFT+DMFT approach with $U_{dp}$ and $U_{pp}$ interactions at $\beta=10$~eV$^{-1}$. The notation of the orbitals is the same as in the Table \ref{DD1}.}
\label{DD3}
\end{table*}

\vspace{0.5cm}

\end{appendix}

\vspace{-1cm}
\end{widetext}


\end{document}